\documentstyle[12pt,fleqn, psfig]{article}
\makeatletter                                               
\def\@eqnnum{{\normalfont \normalcolor [\theequation]}}
\makeatother

\makeatletter
\def\@cite#1#2{({#1\if@tempswa , #2\fi})}
\makeatother

\makeatletter
\def\@biblabel#1{#1.}
\makeatother
\begin{document}
\title{On the thermocapillary motion of deformable droplets  } 

\author{ V.Berejnov\\
\small{Department of Chemical Engineering, Technion, Haifa 32000, Israel}}
\maketitle

%\vskip+3cm
%{\normalsize \bf Running title :} thermocapillary motion of deformable droplets

%\vskip+3cm
%{\underline {For correspondence}}

%{\normalsize \bf e-Mail:} ceranvb@techunix.technion.ac.il
%\bigskip

%{\normalsize \bf TEL:} 972-4-8293562
%\bigskip

%{\normalsize \bf FAX:} 972-4-8230476
%\bigskip

%\newpage
\medskip
\begin{abstract}{In studies on Marangoni type motion of particles the surface tension is often 
approximated as a linear function of temperature.  For deformable particles in a linear external
 temperature gradient far from the reference point this approximation yields 
a negative surface tension which is physically unrealistic. 
It is shown  that H. Zhou and R. H. Davis ({\it J. Colloid Interface Sci.}, {\bf 181}, 60, (1996))
presented calculation where the leading deformable drop moved into a region 
of negative surface tension.
With respect numerical studies the restriction of the migration of two deformable drops is given 
in terms of the drift time.}
\end{abstract}
\bigskip
\bigskip
\bigskip

%{\it \bf Key Words:} surface tension, droplets interaction; thermocapillary migration, deformable droplet.

%\bigskip
%\bigskip
%\bigskip

The bulk fluid motion induced by an interface has been studied for over a century. 
One of the  most interesting  phenomena is the  capillary motion of particles  
through a viscous fluid. 
Young, Goldstein and Block \cite{Young59} and later Bratukhin \cite{Brat75} performed
the first systematic study of the migration of bubbles and droplets.
As noted in  review \cite{Subr92} the capillary motion arises due to gradient
of the surface tension $\gamma$ at the interface as a result of a non--uniform
temperature or surfactant distribution in the surrounding media. 
The surface tension gradient results in a tangential stress on the interface
which causes the motion of the surrounding liquid by viscous traction. 
Then, the droplet or bubble will move in the direction of decreasing interfacial 
tension. It is necessary to note  that the normal component of the capillary forces 
arising during the motion may deform the shape of a particle \cite{Brat75}. 
Young, Goldstein and Block \cite{Young59} and others have shown  that in the limit of high
surface tension (undeformed spherical particle) its motion is controlled by surface
 tension  gradients only. 
Note that the motion of a deformable particle also depends on the surface tension itself.

If the particle moves with constant velocity the transformation of a laboratory coordinate
 system to a coordinate system moving with the particle frame will essentially simplify the solution \cite{Brat75}, \cite{Subr92}, \cite{Ant86}.
Let us denote the particle coordinate system  moving with the droplet velocity  
${\bf U}$ by $O'$ and the laboratory coordinate system by $O$ respectively.
We consider  the coordinate  transform from $O$ to $O'$ 
in the case of a drop moving in the uniform external temperature gradient 
$A{\bf e}_x$ \cite{Brat75}, see Fig.1. For an arbitrary point $F$ we obtain,
\begin{equation}
{\bf R} = {\bf R}^{'} + {\bf U}\:t,\;\;\; {\bf V}_{i}({\bf R},t)={\bf V}_{i}^{'}
({\bf R}^{'})+{\bf U},\;\;\;T_{i}({\bf R},t)=T_{i}^{'}({\bf R}^{'})+ A\:U\:t,
\label{e1}
\end{equation}
where $i=1,2$ correspond to the inner and outer liquid phase, respectively, 
${\bf V}$ is the fluid velocity, $T'$ denotes the 
difference between the temperature $T$ in $O$ and a undisturbed temperature $A U t$
at the center of $O'$, ${\bf R}$ is a radius vector 
which points from $O$ to $F$ and $t$ is the time.

In the limit of an infinitely large surface tension the normal stress boundary condition is not 
modified under the above transformation [\ref{e1}].
However, in the case of finite surface tension, 
this boundary condition requires special attention.
Usually, $\gamma$ is assumed to be linearly dependent on temperature or on concentration is linearized \cite{Adams76},
\begin{equation}
\gamma({\bf R},T) = \gamma_{0}(T_{0}) + 
\left.\frac{\partial\gamma}{\partial T}\right|_{T=T_0} ( T({\bf R}) -
 T_{0}),
\label{gamma}
\end{equation} 
where ${\partial\gamma}/{\partial T}$ is a constant and $T_0$ and $\gamma_{0}$ correspond 
to the reference values of temperature and surface tension, respectively.
Note that for many cases ${\partial\gamma}/{\partial T} < 0$. Due to the transformation 
of $T$ the surface tension $\gamma({\bf R},T)$ is also transformed in the moving 
coordinate system,
\begin{equation}
\gamma^{'}({\bf R}^{'},T^{'}) =  \gamma_{0}(T_{0}) - 
\frac{\partial\gamma}{\partial T}\:A\:{ U}\:t + 
\frac{\partial\gamma}{\partial T}\left( T^{'}({\bf R}^{'}) - 
T_0\right).
\label{tm}
\end{equation} 
The surface tension $\gamma^{'}$ is  time dependent now.
Recall that the surface tension must be positive \cite{Adams76},
\begin{equation}
\gamma \geq 0,\;\;\;\gamma^{'} \geq 0.
\label{ineq}
\end{equation}
From [\ref{tm}] and [\ref{ineq}] follows an upper bound of the drift distance $Ut$ in system $O$ or an upper bound of the time of particle migration in the moving system $O'$.

Ignoring the above restrictions results in the appearance of a negative surface tension
in the course of the particle migration in finite time and thus may lead to a physically
unrealistic behavior of the particle. This restriction is relaxed in the case of the undeformed
drop \cite{Young59}, \cite{Subr81} and \cite{Subr92} where  the normal stress boundary condition
is always satisfied. However, this is not true when the surface tension has a finite value.
We noted that in the literature on thermocapillary migration of drops and bubbles no attention
was paid to this point.  
For example, Zhou and Davis \cite{ZhouD96} first considered 
the problem of axisymmetric thermocapillary migration of {\it  two deformable viscous drops} .
The authors assumed a linear dependence of surface tension on 
temperature. In terms of \cite{ZhouD96} we have
\begin{equation}
\gamma({\bf x}_{s}) = \gamma_{0} + \frac{\partial\gamma}{\partial T}\left( T({\bf x}_{s}) -
 T_{0}({\bf x}_{r})\right) 
\label{ZD1},
\end{equation}
where $T({\bf x}_{s})$ is the temperature at a point ${\bf x}_s$ on the interface 
and $T_{0}({\bf x}_{r})$ is a reference temperature. 
In an attempt to obtain a solution which is independent of the choice of 
${\bf x}_{r}$,  Zhou and Davis fix ${\bf x}_{r}$ to be the intersection point of the axis of symmetry with
the surface of the leading drop, see Fig.1 and their Fig.1 in \cite{ZhouD96}. 
It is important to note that 
this choice of ${\bf x}_{r}$ means a {\it coordinate transform}
from the laboratory frame to the coordinate system moving with the leading droplet.
Hence, the normal stress balance is modified. The other boundary conditions and 
the governing equations remain the same due to the linearity of Stokes and Laplace equations 
\cite{ZhouD96}. For more details see \cite{Brat75} and \cite{Subr81}. 
Zhou and Davis \cite{ZhouD96} give for the dimensionless surface tension in the moving coordinate 
system
\begin{equation}
\bar \gamma({\bf x}_{s}) = 1 - q \:\bar T({\bf x}_{s}),
\label{ZD2}
\end{equation}
where $\bar \gamma=\gamma / \gamma_{0} $ is the 
dimensionless surface tension, $q=aA(-\partial \gamma / \partial T)$ is the dimensionless 
rate of change of the  interfacial tension  due to temperature variation, 
${\bar T} ({\bf x}_{s})= (T({\bf x}_{s}) - T_0({\bf x}_r))/ (a A)$ is the dimensionless 
temperature difference and  $a$  is the radius of the first drop.
It can  readily be seen that Eq.[\ref{ZD2}] defines surface tension which is
positive for any time or migration distances. As we showed before, 
the correct transformation of the linear approximation [\ref{ZD1}] leads to a negative surface 
tension in finite time.
The previous conclusion that physically acceptable solutions must be 
restricted by migration time contradicts Eq. [\ref{ZD2}].

Let us derive the correct form of the transformed surface tension in terms of \cite{ZhouD96}. 
The problem of the migration of two droplets is evolutionary and 
it must be accomplished by a kinematic condition applied on the droplets' surfaces.
The transformation from the laboratory coordinate system to the particle  coordinate
system  are given by \cite{Ant86}:
\begin{eqnarray}
{\bf R} &=& {\bf R}^{'} + \int_{t_1}^{t_2}{\bf U}(t)\:dt,\;\;\; {\bf V}_{i}({\bf R},t)={\bf V}_{i}^{'}
({\bf R}^{'})+{{\bf U}(t)}, \\
T_{i}({\bf R},t)&=&T_{i}^{'}({\bf R}^{'})+ A\:\int_{t_1}^{t_2}{\bf U}(t)\:dt.
\label{e1}
\end{eqnarray}
The migration velocity of the droplet now depends on time and therefore the migration 
distance $U t$ on the right hand side of [\ref{tm}] is given as an integral term,
\begin{equation}
\gamma^{'}({\bf R}^{'},T^{'}) =  \gamma_{0}(T_{0}) - 
\frac{\partial\gamma}{\partial T}A\int_{t_1}^{t_2}{\bf U}(t)\:dt + 
\frac{\partial\gamma}{\partial T}\left( T^{'}({\bf R}^{'}) - 
T_0\right).
\end{equation} 
In terms of \cite{ZhouD96} we have for  the dimensionless surface tension
\begin{equation}
\bar \gamma({\bf x}_{s}) = 1 + \frac{q}{a}\int_{}^{}{\bf U}(t)\:dt - 
q \:\bar T({\bf x}_{s}).
\label{dgam}
\end{equation}
The integral term in Eq. [\ref{dgam}] changes the scenario of a 
numerical calculation. The surface tension changes with  
time and it is necessary to keep $\bar \gamma$ positive. 

We shall now proceed to estimate the time when the 
surface tension of some point ${\bf x}_r$ on the leading drop will 
not satisfy [\ref{ineq}]. For  simplicity let us stay 
in the  laboratory coordinate system, for $\bar \gamma=0$  we obtain 
the relation \footnote{On physical grounds this limit  corresponds to phase transition.}
\begin{equation}
q \:\bar T({\bf x}_{s})=1,
\end{equation}
where $ {\bar T}({\bf x}_{s})=(T({\bf x}_{s}) - 
T_0({\bf x}_0))/ (a A)$ and ${\bf x}_0$ is a reference  point in  system $O$. 
It is readily seen  that the dimensionless length of a spatial frame is  given by $X=\frac{1}{q}$.
Then the maximum transformation distance of the leading drop  
is the  difference between $X$  and the initial position ${\bf x}_r$.
For the case of equal material parameters considered by \cite{ZhouD96} we have 
$a=1$ and  the surface separation distance on  the axes is $\sim 1$.  
Hence the length of the drops' drift is also $\sim 1$. 
Let us assume that the lower bound of the velocities for moving deformable drops is the velocity
of non--deformable drops.
For slightly unequal drops and a large separation distance between their 
centers the velocities are nearly the same  and equal to the Young--Bratukhin value of 0.133..
Following  Eq. [12] in \cite{ZhouD96} we normalize this value with $2/15$ because for inner and outer liquids the
viscosity and the thermal diffusivity are equal.
From this normalization procedure we obtain that the migration velocities are $\sim 1$. 
As a result the critical value of the  migration time is $\sim 1$. 

We developed a  numerical code for solving the problem of the motion 
of two deformable viscous drops in an external temperature gradient \cite{B01}.
Restrictions [\ref{ineq}] were considered  
in the laboratory coordinate system $O$.  In Fig.2 we plotted the evolution of the 
minimum separation distance $d$ between the droplets' surfaces in time. We chose 
$a=1$, $\alpha=0.5$, $q=0.2$ in terms of \cite{ZhouD96}, where $\alpha$ is the droplets radii ratio. 
The dotted curve confines the physical region where [\ref{ineq}] is satisfied.
The curves $1-10$ correspond to different initial separations.
Curve $2$ is in agreement with the results given by Zhou and Davis (see Fig.4 in \cite{ZhouD96}) 
and with the asymptotics for the non--deformable drops \cite{KeChe90}.

Our computations show that the patterns of drops deformations are similar to
those described by \cite{ZhouD96} but correspond to smaller separation distances $d$.
Note that our analysis is restricted by [\ref{ineq}] while the results of \cite{ZhouD96}
lie in the physically unrealistic region.
For initially spherical drops and an initial separation distance $d=0.01$ Fig.3 depicts the series of drops' profiles corresponding to the points $a$, $b$ and $c$ in Fig.2.

\subsection*{ }
The author wish to thank A. M. Leshansky and T. Loimer for helpful discussions.

%\newpage

%\newpage
%{\large \bf Figure legends}
%\bigskip
%\begin{enumerate}
%\item[{\bf Figure 1.}] Geometric sketch of a drop immersed in an
%external temperature gradient $\nabla T$ parallel to its axis of symmetry.

%\item[{\bf Figure 2.}] Evolution of the separation
%distances for two deformable viscous drops under a linear external temperature gradient
%as a function of the initial separation \cite{B01}. The values of the parameters are the same as 
%in \cite{ZhouD96}: $a=1$, $\alpha=0.5$, $q=0.2$.

%\item[{\bf Figure 3.}] Deformation patterns for the initial separation 0.01 and 
%$a=1$, $\alpha=0.5$, $q=0.2$. Figures  $a, b, c$ correspond to the respective points on curve 10 of
%the Fig.2.

%\end{enumerate}

\newpage
\begin{center}
{\large \bf Figures}
\end{center}
\begin{figure}[h!]
\vskip+5cm
\centerline{
\begin{tabular}{cc}
\psfig{figure=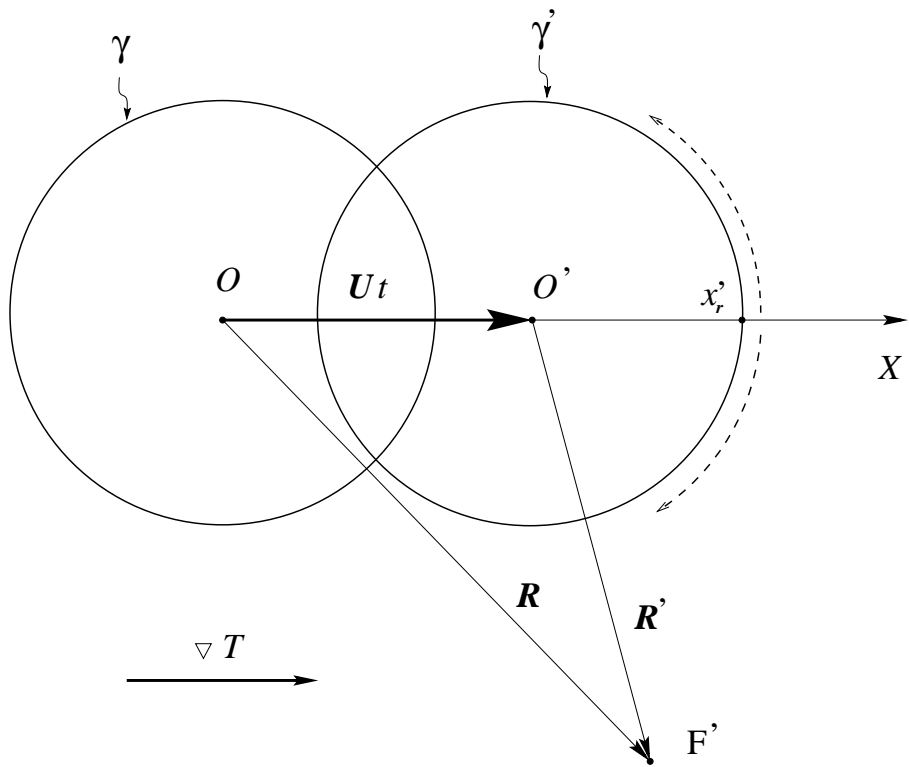,width=12cm}
\end{tabular}}
\caption{
Geometric sketch of a drop immersed in an
external temperature gradient $\nabla T$ parallel to its axis of symmetry.
}
\label{f1}
\end{figure}

\begin{figure}[h!]
\centerline{
\begin{tabular}{cc}
\psfig{figure=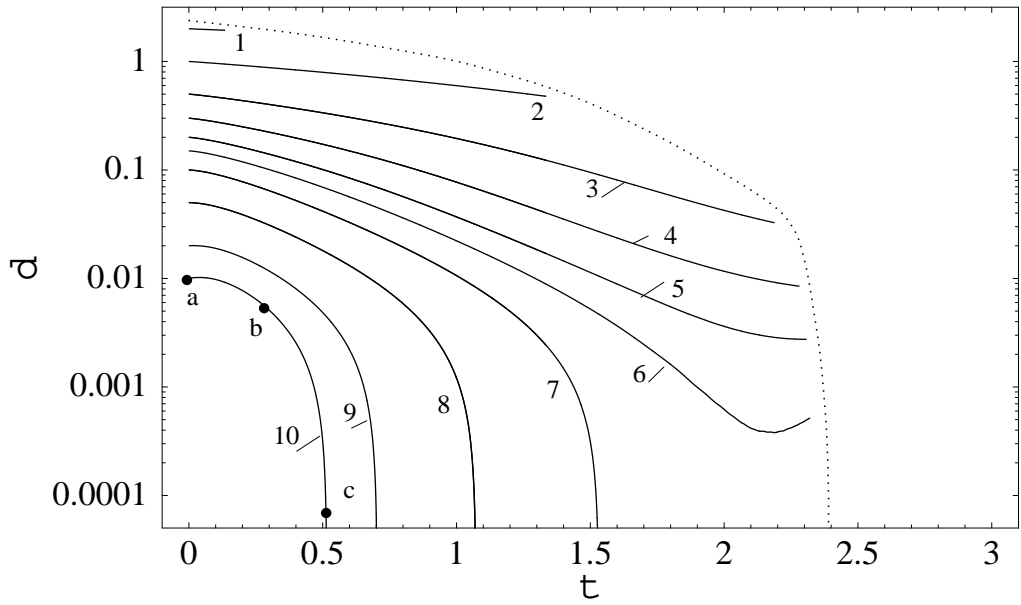,width=15cm}
\end{tabular}}
\caption{
Evolution of the separation
distances for two deformable viscous drops under a linear external temperature gradient
as a function of the initial separation \cite{B01}. The values of the parameters are the same as 
in \cite{ZhouD96}: $a=1$, $\alpha=0.5$, $q=0.2$.
}
\label{f2}
\end{figure}

\clearpage
\begin{figure}[h!]
\vskip+1cm
\centerline{
\begin{tabular}{cc}
\psfig{figure=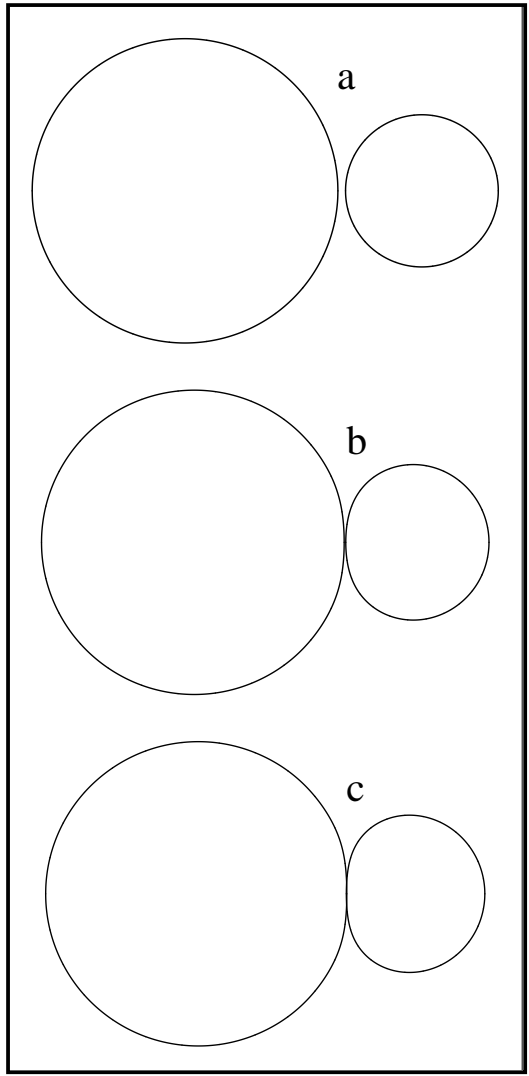,width=7cm}
\end{tabular}}
\caption{ 
Deformation patterns for the initial separation 0.01 and 
$a=1$, $\alpha=0.5$, $q=0.2$. Figures  $a, b, c$ correspond to the respective points on curve 10 of
 the Fig.2.
}
\label{f3}
\end{figure}

\end{document}